\documentclass[aps,prl,10pt,twocolumn,superscriptaddress,balancelastpage,showpacs,reprint,english]{revtex4-1}
\usepackage[T1]{fontenc}
\usepackage[latin9]{inputenc}
\usepackage{amsmath}
\usepackage{amssymb}
\usepackage{graphicx}
\usepackage{color}

\makeatletter

\newcommand{\qq}{\begin{eqnarray}}
\newcommand{\qqq}{\end{eqnarray}}

\newcommand{\p}{\partial}

\newcommand{\bfx}{\mathbf{x}}

\newcommand{\bfr}{\mathbf{r}}

\newcommand{\bfJ}{{\bf J}}
\newcommand{\bfq}{{\bf q}}

\pdfpageheight\paperheight
\pdfpagewidth\paperwidth

\makeatother

\usepackage{babel}

\begin{document}

\title{Capillary interfacial tension in active phase separation}
\author{G. Fausti}
\affiliation{Service de Physique de l'Etat Condens\'e, CEA, CNRS Universit\'e Paris-Saclay, CEA-Saclay, 91191 Gif-sur-Yvette, France}

\author{E. Tjhung}
\affiliation{Department of Physics, University of Durham, Science Laboratories, South Road, Durham DH1 3LE, UK}

\author{M. E. Cates} 
\affiliation{DAMTP, Centre for Mathematical Sciences, University of Cambridge, Wilberforce Road, Cambridge CB3 0WA, UK}

\author{C. Nardini}
\affiliation{Service de Physique de l'Etat Condens\'e, CEA, CNRS Universit\'e Paris-Saclay, CEA-Saclay, 91191 Gif-sur-Yvette, France}
\affiliation{Sorbonne Universit\'e, CNRS, Laboratoire de Physique Th\'eorique de la Mati\`ere Condens\'ee, 75005 Paris, France}

\date{\today}

\begin{abstract}
In passive fluid-fluid phase separation, a single interfacial tension sets both the capillary fluctuations of the interface and the rate of Ostwald ripening. We show that these phenomena are governed by two different tensions in active systems, and compute the capillary tension $\sigma_{\rm cw}$ which sets the relaxation rate of interfacial fluctuations in accordance with capillary wave theory. We discover that strong enough activity can cause negative $\sigma_{\rm cw}$. In this regime, depending on the global composition, the system self-organizes, either into a microphase-separated state in which coalescence is highly inhibited, or into an `active foam' state. Our results are obtained for Active Model B+, a minimal continuum model which, although generic, admits significant analytical  progress.
\end{abstract}


\maketitle
Active particles extract energy from the environment and dissipate it to self-propel~\cite{ramaswamy2017active,Marchetti2013RMP}. Among their notable self-organizing features is phase separation into dense (liquid) and dilute (vapor) regions, even for purely repulsive particles~\cite{Tailleur:08,Cates:15,Fily:12}. Although generically a far-from-equilibrium effect, active phase separation was first described via an approximate mapping onto equilibrium liquid-vapor phase separation~\cite{Tailleur:08,Cates:15}, leading to early speculation that time reversal symmetry might be restored macroscopically in steady state \cite{Speck2014PRL,Tailleur:08,fodor2016far,Brader:15,Maggi:15,szamel2016theory,nardini2017entropy}. Indeed, activity is an irrelevant perturbation near the liquid-vapor critical point, albeit without causing emergent reversibility~\cite{caballero2020stealth}.

Recently it has become clear, however, that bulk phase separation in active systems displays strongly non-equilibrium features. Bubbly phase separation~\cite{tjhung2018cluster} was evidenced in simulations of repulsive self-propelled particles~\cite{stenhammar2014phase,caporusso2020micro}: here large liquid droplets contain a population of mesoscopic vapor bubbles that are continuously created in the bulk, coarsen, and are ejected into the exterior vapor, creating a circulating phase-space current in the steady state. Microphase separation of vapor bubbles~\cite{shi2020self,caporusso2020micro} has been further observed numerically, alongside a similar phase of finite dense clusters, often found in experiments with self-propelled colloids~\cite{Palacci:12,Speck:13} and bacteria~\cite{thutupalli2017boundaries}. Recently, even more intriguing forms of phase separation have been reported in an active system of nematodes, comprising a phase where dense filaments continuously break up and reconnect~\cite{demir2020dynamics}.

Much understanding of active phase separation has been gained from continuum field theories. In the simplest setting~\cite{Wittkowski14,tjhung2018cluster,thomsen2021periodic}, these only retain the evolution of the density field $\phi$, while hydrodynamic~\cite{tiribocchi2015active,singh2019hydrodynamically} or polar~\cite{tjhung2012spontaneous} fields can be added if the phenomenology requires. Their construction, via conservation laws and symmetry arguments, follows a path first introduced with Model B for passive phase separation~\cite{hohenberg1977theory,chaikin2000principles,bray2001interface}. Yet, these field theories differ from Model B because locally broken time-reversal symmetry implies that new non-linear terms are allowed. The ensuing minimal theory, Active Model B+ (AMB+)~\cite{nardini2017entropy,tjhung2018cluster}, including all terms that break detailed balance up to order $\mathcal{O}(\nabla^4\phi^2)$ in a gradient expansion~\cite{nardini2017entropy,tjhung2018cluster}, is defined by 
\qq
\p_t\phi&=&-\nabla\cdot\left(\mathbf{J}+\sqrt{2DM}\mathbf{\Lambda}\right)\,\label{eq:AMB+}\\
 {\bf J}/M &=&-\nabla \mu_\lambda  + \zeta (\nabla^2\phi)\nabla\phi\,\label{eq:AMB+J}\\
\mu_\lambda[\phi] &=& \frac{\delta \mathcal{F}}{\delta\phi} +\lambda|\nabla\phi|^2\label{eq:AMB+mu}
\qqq
where $\mathcal{F} = \int d\bfr \,\left[f(\phi) +\frac{K(\phi)}{2}|\nabla\phi|^2\right]$, $f(\phi)$ is a double-well local free energy,
and $\mathbf{\Lambda}$ is a vector of zero-mean, unit-variance, Gaussian white noises. Model B is recovered at vanishing activity
($\lambda=\zeta=0$), unit mobility ($M=1$) and constant noise level $D$~\cite{hohenberg1977theory}. 

It is known that at low activity (small $\lambda,\zeta$), AMB+ undergoes conventional bulk phase separation. At higher activity, Ostwald ripening~\cite{Bray}, the classical diffusive pathway to macroscopic phase separation, can go into reverse~\cite{tjhung2018cluster}. This explains the emergence of bubbly phase separation and microphase-separated vapor bubbles. (These phases arise when $\zeta,\lambda > 0$; for $\zeta,\lambda < 0$ the identities of liquid and vapor phases are interchanged.) More specific mechanisms, due to hydrodynamics~\cite{matas2014hydrodynamic,singh2019hydrodynamically} or chemotaxis~\cite{liebchen2015clustering,saha2014clusters}, can also piecewise explain some of these phases. AMB+ does not refute such specific explanations but offers a minimal framework to address generic features of active phase equilibria. Its simplicity admits both significant analytical progress, and efficient numerics. 

For active systems showing bulk liquid-vapor phase separation it has been debated, on the basis of numerical and analytical studies, how to define the liquid-vapor interfacial tension~\cite{solon2018generalized2,bialke2015negative,zakine2020surface,lee2017interface,patch2018curvature,omar2020microscopic,hermann2019non}. One key result of this Letter is to confirm that no unique definition is possible. Inspired by work on equilibrium interfaces~\cite{bray2001interface}, we derive an effective equation for the interface height, and calculate the capillary tension $\sigma_{\rm cw}$ which sets the spectrum of capillary waves and the relaxation times of height fluctuations. We find $\sigma_{\rm cw}$ differs from $\sigma$, the tension introduced in~\cite{tjhung2018cluster} to describe the Ostwald process. Whereas $\sigma < 0$ in the reverse Ostwald regime, this does not ensure capillary instability, which instead requires $\sigma_{\rm cw}<0$.
When the latter holds, depending on the global density, we find two new types of active phase separation (Fig.~\ref{fig:e1}), driven by an interfacial instability of Mullins-Sekerka type~\cite{langer1980instabilities}: a microphase-separated droplet state, where coalescence among droplets is highly inhibited, and an `active foam' state.  

As is standard~\cite{hohenberg1977theory,tjhung2018cluster} we  now set $M=1$, assume constant $D,K$, and select $f(\phi)=a(-\phi^2/2+\phi^4/4)$ with $a>0$. (Our results can be extended to any double-well $f$ and any $K(\phi)>0$.)  We set $\zeta>0$, meaning that reversed Ostwald ripening happens only for vapor bubbles. The corresponding results for $\zeta <0$ follow from the invariance of our model under $(\phi,\lambda,\zeta)\to -(\phi,\lambda,\zeta)$. 
We denote by $\phi_1$ and $\phi_2$ the coexisting vapor and liquid densities in the mean-field limit, $D=0$; note that $\phi_{1,2} = \pm 1$ in the passive case only. More generally they are found by changing variables from $\phi$ and $f$ to $\psi$ and $g$: these `pseudo-variables', introduced in~\cite{solon2018generalized} for $\zeta=0$ and then generalised to AMB+~\cite{tjhung2018cluster}, solve $K\p^2 \psi/\p\phi^2 = (\zeta-2\lambda)\p \psi/\p\phi$ and $\p g/\p \psi = \p f/\p \phi \equiv \mu$, whence $\psi =K \left(\exp[(\zeta-2\lambda) \phi/K] -1\right)/(\zeta-2\lambda)$. 
In terms of them, the equilibrium conditions $\mu_1=\mu_2$ and $(\mu\psi-g)_1 = (\mu\psi-g)_2$ which select the binodals $\phi_{1,2}$ still hold~\cite{tjhung2018cluster,solon2018generalized}. (This change of variables is primarily a mathematical device for constructing the phase equilibria; $\psi$ and $g$ have no direct physical significance beyond this.)  All our analytic results are valid in dimensions $d\geq 2$, while our numerics were done in $d=2$ with periodic boundary conditions and system size $L_x\times L_y$, using a pseudo-spectral algorithm with Euler updating~\cite{supp}. 

\begin{figure}[h]\center
  \centering
    \includegraphics[width=1\linewidth]{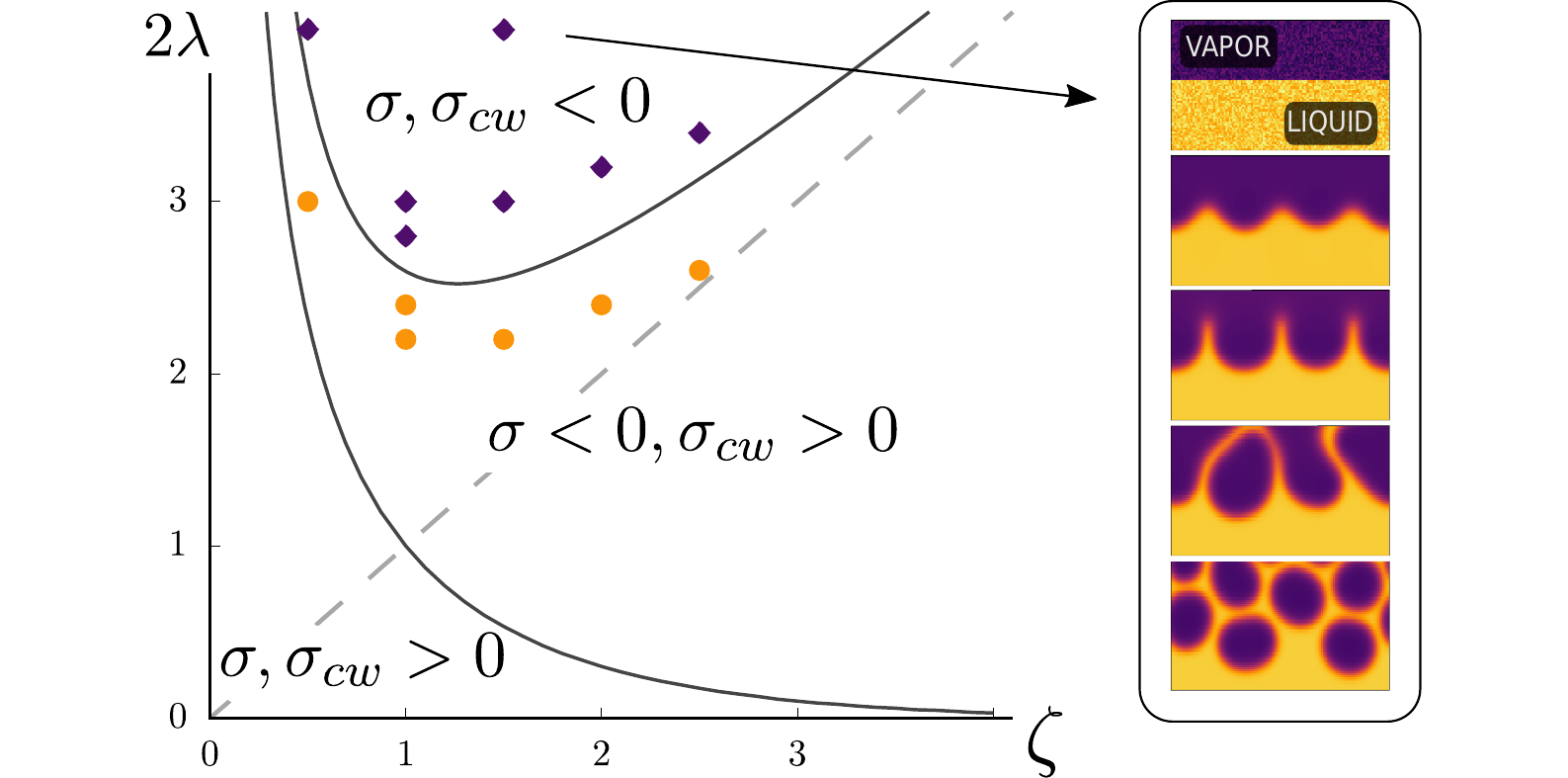}
    \caption{ Mean-field phase diagram for $\zeta>0$, showing sign regimes of interfacial tensions $\sigma$ and $\sigma_{\rm cw}$. 
 When $\sigma_{\rm cw}>0$, the interface is stable and unstable otherwise. Orange circles and blue squares respectively denote the results of direct simulations of AMB+ where the instability of the interface is or is not observed. 
    Right: interfacial instability ($\zeta=1.5, \lambda=2$).}
    \label{fig:e}
\end{figure}

We start, following~\cite{Bray}, by deriving the effective dynamics for small fluctuations of the interface height $\hat{h}(\bfx,t)$ above a $(d-1)$ plane, with in-plane and vertical coordinates $({\bf x},y) = {\bf r}$. We assume the absence of overhangs. On a rapid time-scale, we expect diffusion to quasi-statically relax  $\phi({\bf r},t)$ to a value that depend only on the distance to the interface. For small amplitude, long-wavelength perturbations, the vertical direction and the one normal to the interface are equivalent and we thus can assume that:
\qq\label{eq:phi-varphi}
\phi(\bfr,t)=\varphi(y-\hat{h}(\bfx,t))\,
\qqq
where $\varphi$ is the interfacial profile. By mass conservation, the spatial average of $\hat{h}$ is constant; we set $\hat{h} = 0$. It will turn out that $\hat{h}$ solves a non-local equation in space, so we work in terms of its Fourier transform $h(\bfq_x,t)$. We proceed by plugging (\ref{eq:phi-varphi}) into (\ref{eq:AMB+}) and inverting the Laplace operator. 
We multiply $\nabla^{-2}\partial_t \varphi$ by $\p_y\psi$, integrate across the interface, Fourier transform along the $\bfx$-direction, 
and expand in powers of $h$. Denoting $q=|\bfq_x|$, we obtain~\cite{supp}
\qq\label{eq:AMB-effectve-eq-h}
\p_t h &=& -\frac{1}{\tau(q)} h + \chi +\mathcal{O}( q^2 h^2)\,\\
\frac{1}{\tau(q)} &=& \frac{2\sigma_{\rm cw}(q)q^3}{A(q)}\label{eq:AMB-effectve-eq-h-tau}
\qqq
where
\qq\label{eq:Interfacial-tension-AMB+}
\sigma_{\rm cw}(q) 
&=&
\sigma_{\lambda}+
\frac{3\zeta }{4}
\int dy_1 dy_2\,
\frac{(y_1-y_2)}{|y_1-y_2|}
\frac{\psi'(y_1) \varphi'^2(y_2)}{e^{q|y_1-y_2|}}\\
\sigma_\lambda 
&=&
K \int dy\, \varphi'(y)\psi'(y)
\qqq
and $\chi$ is a zero-mean Gaussian noise with correlations $\langle\chi (\bfq_1,t_1) \chi (\bfq_2,t_2)\rangle = C_\chi(q_1) \delta(\bfq_1+\bfq_2)\delta(t_1-t_2)$, with 
\qq\label{eq:effective-noise}
C_\chi(q) = 
4(2\pi)^{d-1}\frac{D B(q)}{A^2(q)}q\,.
\qqq
In (\ref{eq:AMB-effectve-eq-h-tau},\ref{eq:effective-noise}),  $A(q)
\equiv\int dy_1 dy_2\,\psi'(y_1)\varphi'(y_2) \exp(-q |y_1-y_2|)$ and $B(q)\equiv
\int dy_1 dy_2\,
\psi'(y_1)\psi'(y_2)\exp(-q |y_1-y_2|)\,$. 
Note that (\ref{eq:AMB-effectve-eq-h}) omits nonlinear terms, derived in~\cite{supp}, that previously arose in models of conserved surface roughening~\cite{caballero2018strong,sun1989dynamics}.

The effective height equations (\ref{eq:AMB-effectve-eq-h}-\ref{eq:effective-noise}) are the fundamental analytic results of this Letter. For wavelengths much larger than an interfacial width $\xi\sim \xi_{eq} = (K/2a)^{1/2}$,  we can replace $\sigma_{\rm cw}(q), A(q)$ and $B(q)$ with their limiting values as $q\to0$. These, with a slight abuse of notation, are denoted as $\sigma_{\rm cw},A$ and $B$. Explicitly, the resulting capillary-wave tension $\sigma_{\rm cw}$ obeys
\qq\label{eq:elastic_tension_q0}
\sigma_{\rm cw}
=
 \sigma_\lambda
-\frac{3\zeta}{2} \int dy
\left[\psi(y) -\frac{\psi_1+\psi_2}{2}\right]
\varphi'^2(y)
\qqq
where $\psi_{1,2}=\psi(\phi_{1,2})$ are the pseudo-densities at the binodals. As expected, in the equilibrium limit, $\sigma_{\rm cw}$ reduces to the standard interfacial tension $\sigma_{\rm eq} = K \int dy\, \varphi'^2(y)$ $\lambda,\zeta\to0$~\cite{CatesJFM:2018} which governs not only the capillary fluctuation spectrum, but the Laplace pressure and the rate of Ostwald ripening~\cite{bray2001interface,Bray}.
Switching on activity breaks this degeneracy. Indeed the tension determining the rate of Ostwald ripening of a bubble was given in~\cite{tjhung2018cluster} as
$\sigma 
= 
\sigma_\lambda
-\zeta\int dy\, 
\left[\psi -\psi(0)
\right]
\varphi'^2(y)$, 
where $\psi(0)$ is the value of the pseudo-density at the droplet center. Therefore $\sigma$ is in general not equal to $\sigma_{\rm cw}$.

\begin{figure}[ht]
  \includegraphics[width=1\linewidth]{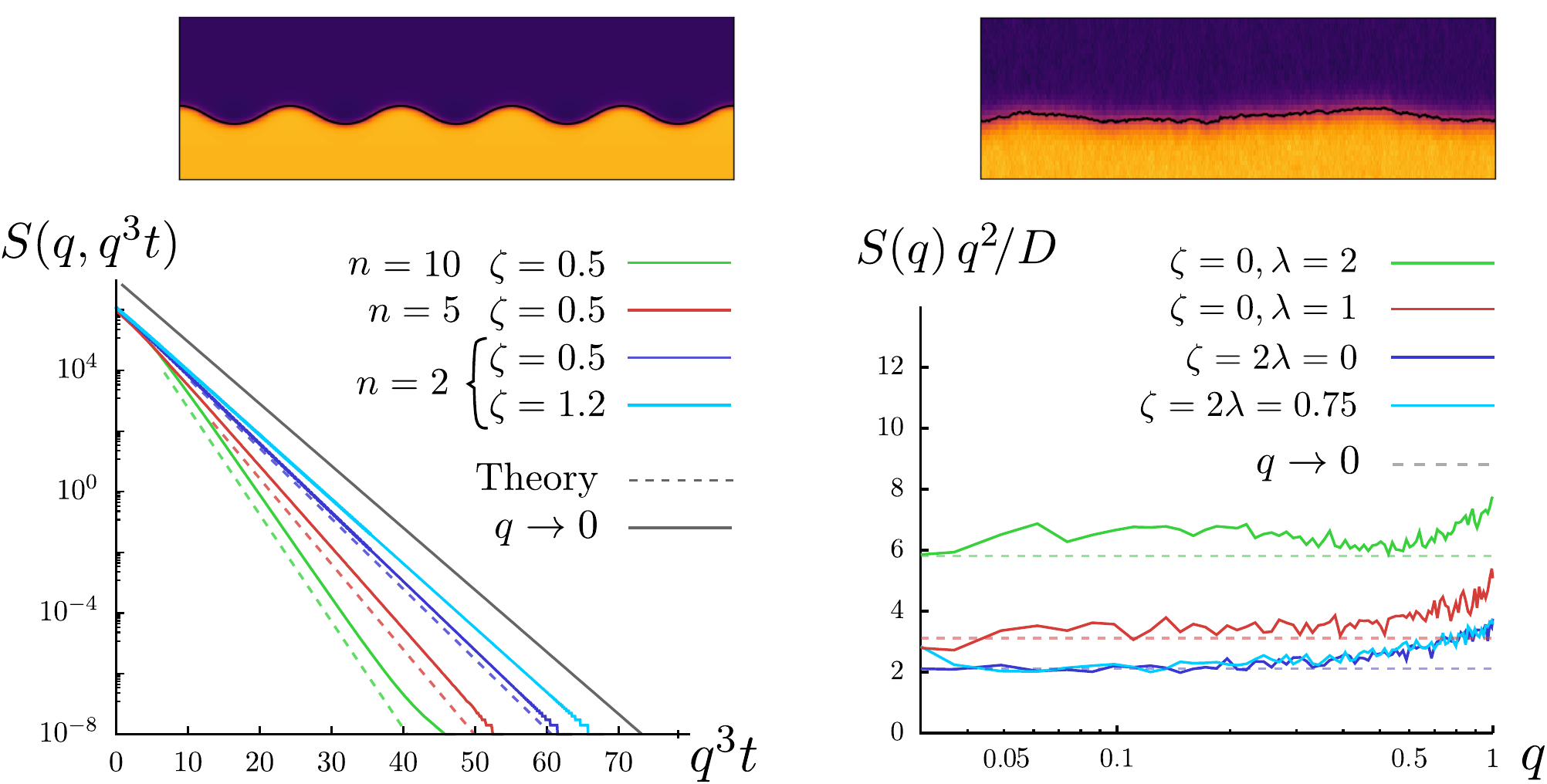}
    \caption{(Left) liquid-vapor interface for $\sigma_{\rm cw}>0$ ($L_x=L_y=256$), $D=0, \zeta=2\lambda$ and its relaxation compared to the theoretical predictions for initial perturbations at wavenumber $2\pi n/L_x$. Dashed lines are  predictions obtained using $\tau(q)$, converging to the $q\to 0$ prediction (continuous line). (Right) snapshot in steady state for $D=5\times 10^{-3}$ and scaled structure factor $q^2 S(q)/D$ vs $q$ compared to the $q\to0$ analytical prediction; results are averaged over $30$ realisations of duration $10^6$ after equilibration.}
    \label{fig:b}
\end{figure}

To gain explicit predictions from (\ref{eq:AMB-effectve-eq-h}-\ref{eq:effective-noise}), we must evaluate $\sigma_{\rm cw}$, $A$ and $B$. This requires knowledge of the interfacial shape $\varphi(y)$. At equilibrium, this is well-known~\cite{CatesJFM:2018}: $\varphi_{\rm eq}(y)=\pm\tanh(y/\xi_{\rm eq})$ with $\xi_{\rm eq}=\sqrt{2K/a}$ and $\sigma_{\rm eq}=\sqrt{8Ka/9}$. (Note that $A = B = 4$ in this case.) Also, whenever $2\lambda=\zeta$ it is readily shown that $\varphi=\varphi_{\rm eq}$ so that $\sigma_{\rm cw}=\sigma_{\rm eq}$, although the Ostwald tensions are $\sigma=\sigma_{\rm eq}(1\mp\zeta/K)$ for bubble growth ($-$) and liquid droplet growth ($+$) respectively \cite{tjhung2018cluster}. We do not have closed-form results for $\sigma_{\rm cw}$ at general $\lambda,\zeta$; however, a change of variable to $w(\varphi)=\varphi'^2$ in the integrals defining $\sigma_{\rm cw},A,B$ allows use of a simple numerical procedure introduced in~\cite{Wittkowski14} and detailed in~\cite{supp} to find the low $q$ behavior. To examine $q\neq 0$ below we instead extract the interface profile from simulations at $D=0$.

Fig.~\ref{fig:e} shows a phase diagram of AMB+ for $\zeta > 0$ at mean-field level, delineating zones of negative $\sigma$ and $\sigma_{\rm cw}$. (There are none at $\zeta>0$ and $\lambda<0$). This provides the full phase diagram of AMB+: the case of $\zeta<0$ follows from Fig.~\ref{fig:e} using the symmetry $(\lambda,\zeta,\phi)\to - (\lambda,\zeta,\phi)$ of AMB+, which interchanges the liquid and vapor identities. 
For small activity, or for $\lambda\zeta<0$, $\sigma_{\rm cw}>0$, even where $\sigma<0$; here vapor bubbles undergoing reversed Ostwald ripening have stable interfaces and, depending on the global density, the system is either micro-phase separated or in bubbly phase separation~\cite{tjhung2018cluster}. At high activity a new regime emerges where $\sigma_{\rm cw}<0$ implying that such interfaces (and also flat ones) become locally unstable.  

We first consider the regime $\sigma_{\rm cw}>0$, where our theory predicts this  capillary tension to govern, via \eqref{eq:AMB-effectve-eq-h-tau}, the relaxation times of the interface $\tau(q)$. To check this, we performed simulations of AMB+ for $D=0$ starting from a phase separated state with the interface perturbed via a single mode; Fig.~\ref{fig:b} confirms that $h(\bfq,t)=h(\bfq,0)\exp(-t/\tau(q))$ as predicted by (\ref{eq:AMB-effectve-eq-h}-\ref{eq:effective-noise}), for either sign of the Ostwald tension $\sigma$. Our theory also predicts the stationary structure factor of the interface $S(q) = \lim_{t\to\infty} \langle |h(\bfq,t)|^2\rangle$:
\qq\label{eq:CWT}
S(q)
=\frac{(2\pi)^{d-1}D}{\sigma_{\rm cw}(q) q^2}\frac{B(q)}{A(q)}
\to_{q \xi^{-1}\ll 1}
\frac{(2\pi)^{d-1} D_{\rm eff}}{\sigma_{\rm cw} q^2}
\qqq
where $ D_{\rm eff} = D(\psi_2-\psi_1)/(\phi_2-\phi_1)$ is an effective capillary temperature. Eq. (\ref{eq:CWT}) generalizes capillary wave theory. Its equilibrium analog, $S(q)\propto D/ \sigma_{\rm eq} q^2$~\cite{rowlinson2013molecular}, is often justified using equipartition arguments but, even in equilibrium, higher order gradient terms give sub-leading corrections at finite $q$~\cite{meunier1987liquid,blokhuis1993van}. Activity impacts the interfacial fluctuations by renormalizing the temperature $ D\to D_{\rm eff}$ and, separately, replacing $\sigma_{\rm eq}$ with $\sigma_{\rm cw}$. Even though (\ref{eq:CWT}) also neglects the additional nonlinearities omitted from (\ref{eq:AMB-effectve-eq-h}), it is quite accurate at small $D$ (Fig.~\ref{fig:b}). The use of capillary wave theory in phase-separated active systems was previously advocated heuristically~\cite{patch2018curvature,bialke2015negative,lee2017interface} but  until now, only qualitative estimates were provided for the coefficient $ D_{\rm eff}/\sigma_{\rm cw}$ in (\ref{eq:CWT}). 

\begin{figure}[h!]\center
  \centering
		\includegraphics[width=.9\linewidth]{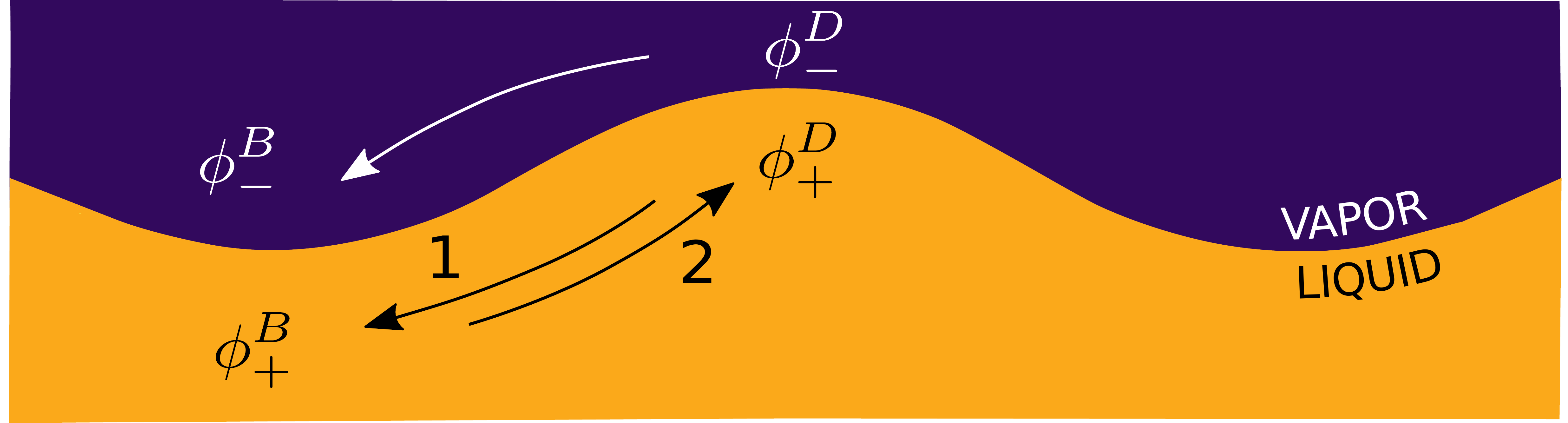}
                 \caption{Instability at $\sigma_{\rm cw}<0$. The densities on the two sides of the interface adjust quasi-statically at values that depend on its local curvature. The ensuing diffusive density fluxes on the vapor side is always stabilising (white arrow); that in the liquid is stabilising when $\sigma>0$ ($\textrm{arrow } 1$) and becomes destabilising when $\sigma<0$ ($\textrm{arrow } 2$). This (one-sided) reverse-Ostwald current does not trigger an instablity unless the current in the liquid outweighs that in the vapor, which requires $\sigma_{\rm cw}<0$.}
    \label{fig:inst-mech}
\end{figure}

\begin{figure*}
  \centering
  \includegraphics[width=1.\linewidth]{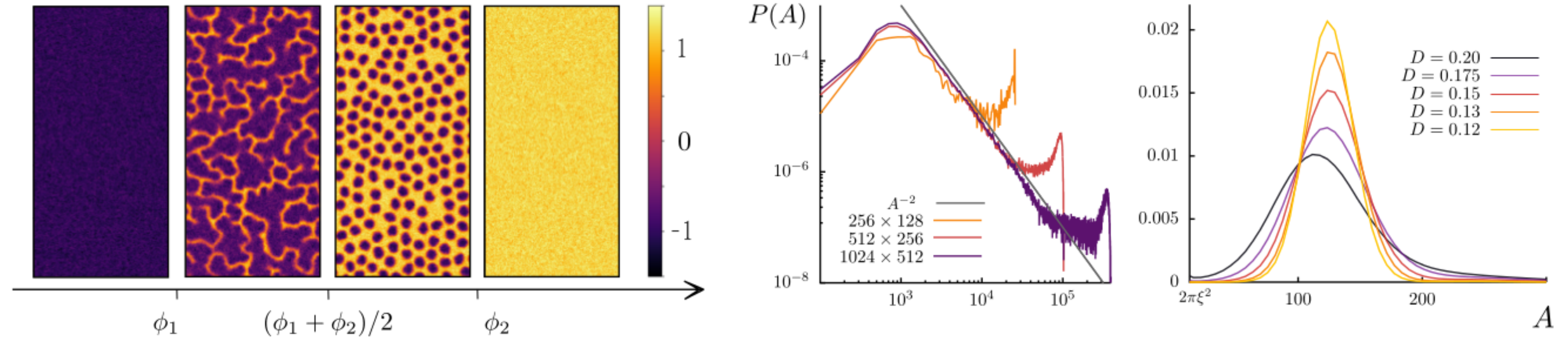}
    \caption{ (Left) phase diagram when $\sigma_{\rm cw}<0$ as a function of the global density $\phi_0=-1, -0.4, 0.4, 1.2$ at  $D=0.05, L_x=256, L_y=512$ and $\lambda=1.75, \zeta=2$, for which $\phi_1=-0.9, \phi_2=1.08$. At high and low $\phi_0$, the system is homogeneous (liquid or vapor states). Within the binodals, when $\phi_0>(\phi_1+\phi_1)/2$, the system shows microphase-separated vapor bubbles whose coalescence is highly inhibited. At lower $\phi_0$, the system forms a continuously evolving active foam state. (Middle and Right): area distribution of vapor regions for the active foam state ($\phi_0=-0.4$) and in the microphase-separated state ($\phi_0=0.2$, noise values in the legend).}
    \label{fig:e1}
\end{figure*}

When $\sigma_{\rm cw}<0$, a drastically new  non-equilibrium phenomenology arises. Although the vapor--liquid interface is unstable to height fluctuations, the system remains phase separated. For, unlike in equilibrium where demixing itself cannot be sustained at negative tension, the active interface does not undergo diffusive collapse but remains linearly stable against normal perturbations $\phi(\bfx,y)=\varphi(y)+\p_y \epsilon(y)$~\cite{supp,shinozaki1993dispersion,bricmont1999stability}. 

Next, we numerically simulated AMB+ at $D=0$, with a noisy initial condition. Orange and blue dots in Fig.~\ref{fig:e} respectively represent cases where the interfacial fluctuation is damped or amplified (Movie 1), showing the accuracy of our analytical predictions. Computing $\tau(q)$ shows that the first unstable mode is at the lowest $q$ available; thus the transition line $\sigma_{\rm cw}=0$ is critical. 

The interfacial instability mechanism (Fig. \ref{fig:inst-mech}) is reminiscent of the Mullins-Sekerka instability in solidification~\cite{langer1980instabilities}. In both cases the instability is driven by a single diffusing field: latent heat in crystal growth, and density here. Such a diffusing field settles to quasi-stationary values $\phi_{\pm}^{B,D}$ on the two sides of the interface which depend on the local curvature. By approximating $\phi_{\pm}^{B,D}$ as the densities near the interface of a vapor bubble (B) or liquid droplet (D), we find that the diffusive current on the vapor side is always stabilizing. In contrast, depending on whether Ostwald ripening is normal or reversed, the current on the liquid side is stabilizing or destabilizing. Reversed Ostwald ripening is however not sufficient to drive overall instability of the interface; this arises only if the current on the liquid side is stronger than the one on the vapor. This condition sets the threshold beyond which $\sigma_{\rm cw}<0$. Measuring the steady state currents confirms this mechanistic picture~\cite{supp}. 
 
We now report simulations with a small but finite noise level to ensure reproducible steady states. Starting from a near-uniform initial state, we find that the final phase separation is strongly affected by interfacial instability. The stable case, $\sigma_{\rm cw}>0$, was explored in \cite{tjhung2018cluster}. For the unstable case, $\sigma_{\rm cw}<0$, the stationary states seen by varying the global density $\phi_0=\int \phi d\bfr/V$ are reported in Fig.~\ref{fig:e1} and Movie 2. 
When $\phi_0$ lies outside the mean-field binodals $\phi_{1,2}$, the system remains homogeneous. Within them, at large $\phi_0$  where the liquid is the majority phase, we find a microphase-separated state where coalescence of crowded bubbles is highly inhibited. The bubble size distribution $P(A)$ is strongly peaked, increasingly so as noise decreases, suggesting that the average bubble size $\langle A\rangle $ is finite when $D\to0$. Our results are converged in time for $D> 0.1$; at lower noise the system gets trapped into metastable states, evolving only because of rare fluctuations of the bubbles interface~\cite{supp}. Clearly, the average size is not set by the most unstable mode of the flat interface, as the steady state is attained through secondary instabilities (Movie 1 and 3). This phenomenology is at odds with the bubble phase at $\sigma_{\rm cw}>0$ \cite{tjhung2018cluster}, where a dynamical balance between nucleation, coalescence and reversed Ostwald causes $\langle A\rangle \to\infty$ when $D\to0$. The difference between these two microphase separated states is also apparent dynamically when starting from bulk phase separation (Movie 3).

When the liquid is the minority phase, bubbles cannot avoid touching and coalescing. One might expect that the system attains a micro-phase separated state of liquid droplets (for $\zeta>0$); this is not the case because,  as is clear from our mechanistic argument above, the interfaces bends toward the vapor side. Instead, we find a distinctive form of phase separation, which we call the `active foam' state. Thin filaments of liquid are dispersed in the vapor phase, which continuously break up and reconnect. This state is previously unknown in active scalar models but resembles patterns that can arise, by a different mechanism, in active liquid crystals~\cite{maryshev2020pattern}. The filaments are bent on the most unstable length-scale of the flat interface. The area distribution of vapor regions (Fig. \ref{fig:e1}b) is now peaked at at size that corresponds to the merging of two bubbles, but a power-law tail $A^{-2}$ emerges, only cut off by the system-size. The boundaries in $\phi_0$ between the different phases of Fig. \ref{fig:e1} are qualitative: while the vapor density is almost independent of $\phi_0$, the liquid density varies~\cite{supp}. 

The techniques introduced here could help elucidate $\sigma_{cw}$ in particle-based active models, by applying them to various field-theoretical descriptions obtained by explicit coarse-graining~\cite{tjhung2018cluster,solon2018generalized,bickmann2020collective}, or to describe confluent biological tissues, where the measured interfacial tension was shown to be dependent on the protocol~\cite{sussman2018soft}. The roughening properties of the interface also merit further study: the anomalous scaling found in particle-based simulations was interpreted to be in the Edwards-Wilkinson universality class~\cite{patch2018curvature,lee2017interface}. Dimensional analysis~\cite{tauber2014critical} of our linear theory instead gives the critical exponents $z=3$ and $\chi=(z-d)/2$, where $\langle \hat{h}(\bfx, t)\hat{h}(\bfx',t)\rangle\sim |\bfx-\bfx'|^{2\chi}$ and $\langle \hat{h}(\bfx, t) \hat{h}(\bfx,t')\rangle\sim |t-t'|^{2\chi/z}$. The impact of non-linearities should be studied by renormalisation methods. 

Finally, it is remarkable that (a) the capillary tension can likewise become negative, and that (b) this leads to new types of phase separation including active foam states. Our generic field-theoretical approach is agnostic as to the microscopic mechanisms underlying activity (and even phase separation). Therefore the microscopic ingredients needed for our new phases remain to be identified. For the same reason, we expect them to be widely present in phase-separating systems with locally broken detailed balance: besides motility-induced phase separation~\cite{Cates:15}, applications might encompass cell sorting in biological tissues~\cite{sussman2018soft}, tumor invasion~\cite{kang2020tumor} and sociophysics~\cite{grauwin2009competition}.

\begin{acknowledgments}
The authors acknowledge H. Chat\'e, A. Patelli and J. Stenhammar for several discussions. GF was supported by the CEA NUMERICS program, which has received funding from the European Union's Horizon 2020 research and innovation program under the Marie Sklodowska-Curie grant agreement No 800945. CN acknowledges the support of an Aide Investissements d'Avenir du LabEx PALM (ANR-10-LABX-0039-PALM). Work funded in part by the European Research Council under the Horizon 2020 Programme, ERC grant agreement number 740269 and by the National Science Foundation under Grant No. NSF PHY-1748958, NIH Grant No. R25GM067110 and the Gordon and Betty Moore Foundation Grant No. 2919.02. MEC is funded by the Royal Society. 
\end{acknowledgments}

\bibliographystyle{apsrev4-1}
\bibliography{biblio_Giordano.bib}

\appendix
\section{Details on the numerical analysis}
We integrated AMB+ with a parallel pseudo-spectral code employing periodic boundary conditions and Euler time-update. In all our simulations we set $K=1$, $a=1/4$, spatial spatial discretisation $\Delta x=1$, time-step $dt=2\times 10^{-2}$; we checked that our results are stable upon decreasing $\Delta x$ and $dt$.

\subsection{Identification of the interface}
To locate the interface, for every $x$ coordinate, we find the first pixel $\bar{y}$ such $\phi(x, \bar{y})>\phi_{th}$, where $\phi_{th}=(\phi_1+\phi_2)/2$. We then place the interface at 
\qq
	h(x) = \bar{y} + \frac{\phi(x, \bar{y})-\phi_{th}}{\phi(x, \bar{y}) - \phi(x, \bar{y}+\Delta x)}\,.
\qqq
Different values of $\phi_{th}$ do not change the results as far as $\phi_{th}$ is sufficiently far from the binodals.
\subsection{ Measure of the bubble-size distribution }
To measure the distribution of bubbles area, we first transformed the density field in a binary matrix using the threshold $\phi_{th}$. Scanning this matrix sequentially, we then applied a  breadth-first search algorithm; the outcome is an $L_x\times L_y$ matrix where each pixel is labeled accordingly to the cluster it belongs to. Building the probability distribution function (PDF) of the vapor regions is then straghtforward. The analysis was performed on a smoothened density field obtained by running a few time-steps of the dynamics without noise.


\section{Effective interface equation in the mean-field approximation}
In this Appendix we detail the derivation of the effective interface equation (5) of the main text. We start from the mean-field problem ($D=0$) in Appendix \ref{app:effective_int_eq_d0} and consider the effect of noise in Appendix \ref{app:noise}. We also obtain the leading non-linear terms, of order $\mathcal{O}(h^2)$, that correct eq. (5).

The full, non-linear, effective equation for $h$ that we obtain is
\qq\label{eq:app:AMB-effectve-eq-h_nonlinear}
\p_t h = &-&\frac{2\sigma_{cw}(q) q^3}{A(q)} h 
+
\zeta \frac{C(q)}{A(q)} \big\{-q^2 \mathcal{F}\left[ |\nabla_{\bfx} \hat{h}|^2 \right]\nonumber\\
&+&\mathcal{F}\left[\nabla_{\bfx}\cdot (\nabla_{\bfx}^2 \hat{h} \nabla_{\bfx} \hat{h})\right]
\big\}
+\chi
+\mathcal{O}(\zeta q^3 h^3)
\qqq
where $q=|\bfq |$,  $\mathcal{F}[\cdot ] = \int d\bfx \,e^{-i \bfq\cdot \bfx}\,\cdot\, $ is the Fourier transform operator along the $\bfx$ direction (the same convention for the Fourier transform is used throughout what follows), and 
\qq
C(q) = \int dy_1 dy_2\, \psi'(y_1) \varphi'^2(y_2) e^{-q |y_1-y_2|}\,.
\qqq
The other quantities appearing in (\ref{eq:app:AMB-effectve-eq-h_nonlinear}) were defined in the main text and are reported here for convenience:
\qq 
&&\sigma_{\rm cw}(q) 
=
\sigma_{\lambda}+
\frac{3\zeta }{4}
\int dy_1 dy_2\,
\frac{(y_1-y_2)}{|y_1-y_2|}
\frac{\psi'(y_1) \varphi'^2(y_2)}{e^{q|y_1-y_2|}}\nonumber \\
&&\sigma_\lambda = K 
\int dy\,\varphi'(y) \psi'(y)\label{app:eq:sigma_lambda} \\
&& A(q) = \int dy_1 dy_2\,\psi'(y_1)\varphi'(y_2) \exp\left(-q |y_1-y_2|\right)\label{app:eq-Aq}\\
&& B(q)=
\int dy_1 dy_2\,
\psi'(y_1)\psi'(y_2)\exp(-q |y_1-y_2|)\,.\label{app:eq-Bq}
\qqq
Finally, the noise $\chi$ is Gaussian and has correlations $\langle\chi (\bfq_1,t_1) \chi (\bfq_2,t_2)\rangle = C_\chi(q_1) \delta(\bfq_1+\bfq_2)\delta(t_1-t_2)$, where 
\qq\label{eq:app:effective-noise}
C_\chi(q) = 
4(2\pi)^{d-1}\frac{D B(q)}{A^2(q)}q\,.
\qqq


\subsection{Effective interface equation for D=0}\label{app:effective_int_eq_d0}
Assuming 
\qq\label{eq:app-Ansatz}
\phi(\bfr,t)=\varphi(y-\hat{h}(\bfx,t))
\qqq
in the AMB+ dynamics, we obtain
\qq\label{eq:h-intermediate-app}
\nabla^{-2}\partial_t \varphi &= 
f'(\varphi) - K \nabla^2\varphi +\lambda|\nabla\varphi|^2\nonumber\\
&-\zeta \nabla^{-2}\nabla\cdot\left[(\nabla^2\varphi)\nabla\varphi\right] 
\qqq
where $\hat{g}(\bfx,y) = \nabla^{-2}\hat{s}(\bfx,y)$ means that $\hat{g}$ solves $\nabla^2 \hat{g} =\hat{s}$. It is easy to show that the Fourier transform of $\hat{g}$ along $\bfx$ is given by
\qq\label{eq:app:inverse-lapl}
g(\bfq,y) = -\frac{1}{2 q}\int dy_1 e^{-q |y-y_1|}s(\bfq,y_1)\,.
\qqq

Let us first consider the equilibrium case $\lambda=\zeta=0$, hence generalizing the approach of~\cite{bray2001interface} to arbitrary $q$-values. Applying the chain rule to (\ref{eq:h-intermediate-app}) gives
\begin{align} \label{eq:A12_def-0}
-\nabla^{-2} \left[   \varphi' \partial_{t} \hat{h}\right] &=\\
&  f'(\varphi) - K \varphi''(1+|\nabla_{\bfx} \hat{h}|^2) + K \varphi'\nabla_{\bfx}^2 \hat{h}\nonumber
\end{align}
where $\nabla_{\bfx}$ is the gradient with respect to $\bfx$.
We then multiply by $\varphi'$ and integrate over $u=y-\hat{h}(\bfx,t)$ across the interface to get
\begin{align} \label{eq:A12_def}
-\int du \varphi'(u) \nabla^{-2} \left[   \varphi'(u) \partial_{t} \hat{h}\right] = \Delta f  + \sigma_{eq}\nabla_{\bfx}^2 \hat{h}
\end{align}
where $\Delta f=f(\phi_2)-f(\phi_1)$, we have assumed that $\phi(y\to\infty)=\phi_2$ and $\phi(y\to-\infty)=\phi_1$, and that $\varphi'$ vanishes in the bulk. Fourier transforming along the $\bfx$ direction and using (\ref{eq:app:inverse-lapl}) gives
\qq\label{eq:app:AMB-effectve-eq-h-1}
\p_t h = -\frac{2\sigma_{eq} q^3}{A_{eq}(q)} h 
\qqq
where $A_{eq}(q)
=\int dy_1 dy_2\,\varphi'(y_1)\varphi'(y_2) \exp\left(-q |y_1-y_2|\right)$. Observe that the term coming from $\Delta f$ in (\ref{eq:A12_def}) is proportional to $q\delta(\bfq)$ and thus vanishes.
Eq. (\ref{eq:app:AMB-effectve-eq-h-1}) is the deterministic part of the effective interface equation for Model B.

We now consider $\lambda, \zeta\neq0$. From (\ref{eq:h-intermediate-app}), the analog to (\ref{eq:A12_def-0}) now reads
\qq\label{eq:A12_def-B+}
-\nabla^{-2} \left[ \varphi'(u) \partial_{t} \hat{h}\right] = \mu_{\lambda} + \mu_{\zeta}
\qqq
where
\qq
\mu_{\lambda} &=& f'(\varphi) +(1+|\nabla_{\bfx} \hat{h}|^2)(\lambda\varphi'^2- K \varphi'') + K\varphi' \nabla_{\bfx}^2 \hat{h} \nonumber\\
\mu_{\zeta} &=& - \zeta \nabla^{-2}\left\{ \nabla_{\bfx}\cdot \Big[
\left(\varphi'' |\nabla_{\bfx} \hat{h}|^2 - \varphi' \nabla_{\bfx}^2 \hat{h} + \varphi''\right)\right.\label{eq:app:mu_zeta}\\
&&\left( -\varphi' \nabla_{\bfx} \hat{h}  \right)
\Big]
 \left. + \partial_y\left[ 
\left(\varphi'' |\nabla_{\bfx} \hat{h}|^2 - \varphi' \nabla_{\bfx}^2 \hat{h}+\varphi'' \right)\varphi'\right]\right\}\,.\nonumber
\qqq

In order to progress we need to introduce the pseudo-variables~\cite{tjhung2018cluster} $\psi,g$, defined as the solution to 
\qq\label{eq:def-psi-g}
\frac{\p^2 \psi}{\p\phi^2} = \frac{\zeta-2\lambda}{K}\frac{\p \psi}{\p\phi} \quad\text{and}\quad \frac{\p g}{\p \psi} = \frac{\p f}{\p \phi} \,
\qqq
such that $\psi\to\phi$ and $g(\phi) \to f(\phi)$ in the passive limit ($\lambda\to0$ and $\zeta\to0$). These quantities play the same technical role as the one played by the density $\phi$ and by the local free energy $f$ in equilibrium systems for computing the binodals~\cite{tjhung2018cluster,solon2018generalized} or for computing the surface tension $\sigma$~\cite{tjhung2018cluster}, although they lack the analogous physical interpretation. 

We then multiply (\ref{eq:A12_def-B+}) by $\psi'$, integrate across the interface and apply the Fourier transform along $\bfx$. For the left hand side of (\ref{eq:A12_def-B+}) we obtain
\qq\label{eq:app:pt-int}
\frac{A(q)}{2 q} \p_t h(\bfq ,t)
\qqq
where $A(q)$ is given in the main text and in (\ref{app:eq-Aq}).

Concerning the right hand side of (\ref{eq:A12_def-B+}), the first term in $\mu_{\lambda}$ becomes
\qq
\delta(\bfq) \int_{-\infty}^{\infty} du\, \psi'(u) f'(\varphi) =
g(\psi_2)-g(\psi_1)=0\,\nonumber
\qqq
where we used the definition of $g$. 
To evaluate the second term in $\mu_\lambda$ we observe that 
\qq\label{eq:app:pt-int-1}
&
\int_{-\infty}^{\infty}du\,
(\lambda\varphi'^2(u)- K \varphi''(u))\psi'(u)\nonumber\\
&=
\frac{\zeta}{2}
\int_{-\infty}^{\infty}du\,
\varphi'^2(u)\psi'(u)
\qqq
where we have used (\ref{eq:def-psi-g}). The contribution in (\ref{eq:app:pt-int-1}) will be cancelled by an opposite one coming from $\mu_\zeta$. The third term in $\mu_\lambda$ gives
\qq\label{eq:app:pt-int-2}
- q^2 \sigma_\lambda h(\bfq,t)
\qqq
where $\sigma_\lambda$ is defined in (\ref{app:eq:sigma_lambda}).

We now consider $\mu_\zeta$ in (\ref{eq:app:mu_zeta}). Expanding in powers of $\hat{h}$, we have
\qq
&\nabla_{\bfx}^2\left(\frac{1}{\zeta}\mu_\zeta\right)
=
-\frac{1}{2} \p^2_y (\varphi'^2) 
+\frac{3}{2} \p_y (\varphi'^2) \nabla^2_{\bfx} \hat{h}\\
&-\varphi'^2 \nabla_{\bfx}\cdot [\nabla_{\bfx}^2 h \nabla_{\bfx}\hat{h}]
- \p^2_y(\varphi'^2) |\nabla_{\bfx}\hat{h}|^2
+\mathcal{O}(\zeta q^3 h^3)\,.
\nonumber
\qqq
We then invert the Laplacian using (\ref{eq:app:inverse-lapl}) and use 
\qq\label{app:eq:puexp}
\p_u e^{-q|y-u|} = q\,\, \textrm{sgn}(y-u) e^{-q |y-u|}
\qqq
where $\textrm{sgn}$ is the sign function. Applying the same procedure as before and adding up the result with (\ref{eq:app:pt-int}), (\ref{eq:app:pt-int-1}), (\ref{eq:app:pt-int-2}) we obtain the deterministic part of (\ref{eq:app:AMB-effectve-eq-h_nonlinear}).

\begin{figure}[ht!]\center
  \centering
        \includegraphics[width=1\linewidth]{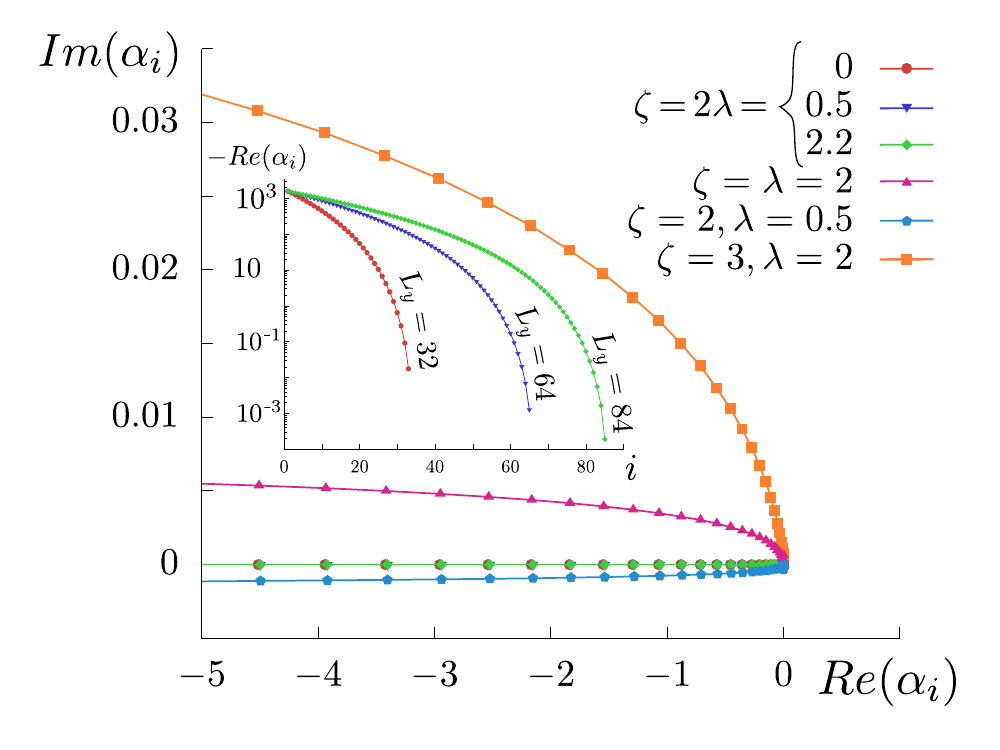}
                 \caption{ Eigenvalues $\alpha_i$ of $\mathcal{L}(q_1,q_2)$ for a system of linear size $L_y=128$ in the vicinity of $Re(\alpha_i)=0$ (the minimal eigenvalues for this system size have real part of order $-10^3$), spatial discretization $\Delta x=1$ and values of the parameters as in the legend. The solid lines are our estimate of the continuous spectrum of $\mathcal{L}(q_1,q_2)$ in the infinite system-size limit, obtained from a fit of the discrete spectrum. Although the spectrum is modified with respect to the equilibrium case, its crucial properties are not: all eigenvalues are negative and touch $0$ in the infinite system-size limit implying that the vapor-liquid interface is stable to normal perturbations but disturbances will only decay algebraically in time. Inset: $-Re(\alpha_i)$ for $\zeta= 2, \lambda=0.5$ as a function of $i$ for three system sizes, showing that the eigenvalue with the largest real part approaches zero.}
    \label{fig:spectrum}
\end{figure}

\subsection{Effect of $D\neq0$}\label{app:noise}
We now consider $D\neq 0$. Our goal is twofold: first, we derive the noise $\chi$ that enters in the effective interface equation (\ref{eq:app:AMB-effectve-eq-h_nonlinear}); second, we show that the Ito term that might correct (\ref{eq:app:AMB-effectve-eq-h_nonlinear}) actually vanishes. We start from this latter point by rewriting (\ref{eq:app-Ansatz}) as 
\qq
\phi(\bfr,t)=\varphi\left(y-\frac{1}{(2\pi)^{d-1}}\int d\bfq \,h(\bfq ,t) e^{i \bfq\cdot \bfx}\right)\,.
\qqq
The time derivative of $\phi$ gives
\qq
\p_t \phi &=
-\varphi' \p_t \hat{h} 
+
\frac{1}{(2\pi)^{2(d-1)}}\varphi'' \int d\bfq_1 d\bfq_2\, e^{i(\bfq_1+\bfq_2)\cdot \bfx} \nonumber\\
&2 D q_1\frac{ B(q_1)}{A^2(q_1)} \delta(\bfq_1+\bfq_2)
\qqq
and hence
\qq
\p_t \phi=
-\varphi' \p_t \hat{h} 
+
\frac{2D}{(2\pi)^{2(d-1)}}\varphi''
\int d\bfq \,\,q\frac{ B(q)}{A^2(q)}\,\label{eq:app-ptphi_Dnot0}
\qqq
where last term in (\ref{eq:app-ptphi_Dnot0}) is the Ito contribution. It is then easy to show that this term gives a contribution proportional to $q \delta(q)$ in $\p_t h$, and hence vanishes.

We are left with deriving the noise $\chi$ and show that its correlation is given by (\ref{eq:app:effective-noise}). We first consider 
	\qq
	\xi(\bfx,y,t)=(\nabla^{-2} \eta)(\bfx,y,t)\,
	\qqq
where $\eta = -\nabla\cdot\sqrt{2D} \mathbf{\Lambda}$. Being a linear transformation of $\eta$, $\xi$ is also a Gaussian noise. Its correlation reads
\qq
C_{\xi}(\bfx,y,t) 
=
-2D \nabla^{-2}\delta(\bfx)\delta(y)\delta(t)\,
\qqq
and its Fourier transform along $\bfx$ 
\qq
C_\xi(q,y,t)
	=
	\frac{D}{q}e^{-q|y|}\delta(t)
\qqq
where we have used (\ref{eq:app:inverse-lapl}). The noise $\chi$ is given by
\qq
\chi(\bfq,t) 
=
\frac{2q}{A(q)}\mathcal{F}\left[
\int du \,
\psi'(u) \xi(\bfx,u+\hat{h}(\bfx,t),t)
\right] (\bfq,t)\nonumber
\qqq
which is also Gaussian.
It is now straightforward to show that the correlation of $\chi$ is given by (\ref{eq:app:effective-noise}). This concludes the derivation of the effective interface equation (\ref{eq:app:AMB-effectve-eq-h_nonlinear}).

\begin{figure*}[ht!]\centering
        \includegraphics[width=0.9\linewidth]{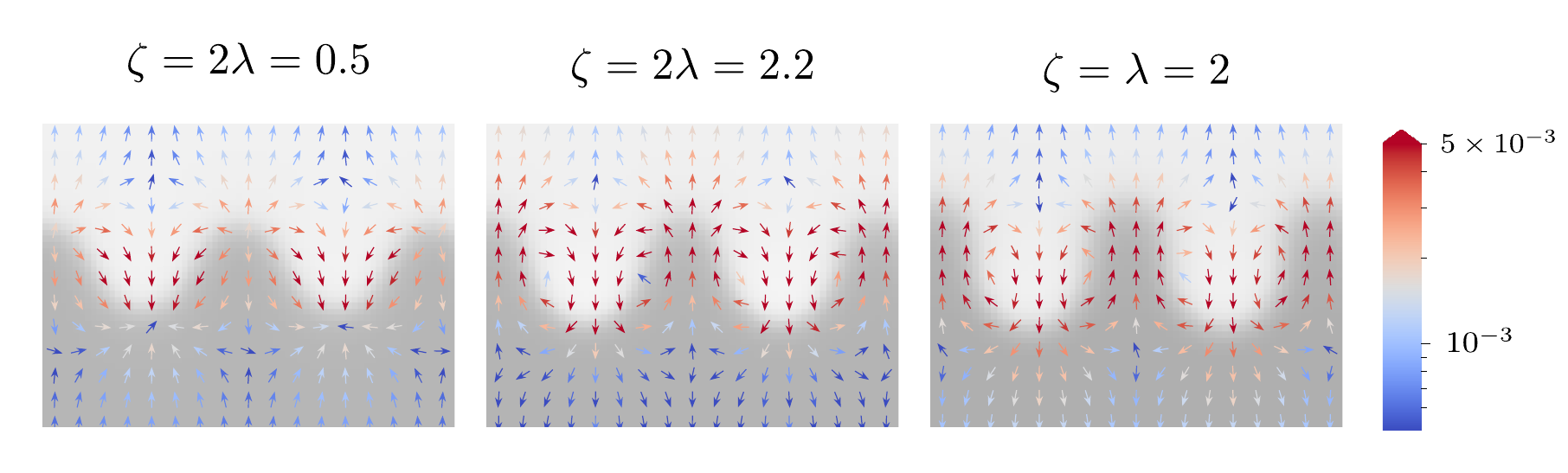}
                 \caption{  Current $\bfJ$ close to (left) the stable interface  and normal Ostwald ripening ($\zeta=2\lambda=0.5$), (middle) stable interface and reversed Ostwald ripening for bubbles ($\zeta=2\lambda=2.2$) and (right) unstable interface  ($\zeta=\lambda=2$). Simulations are at mean-field ($D=0$). The liquid is shown in dark gray and the magnitude of the current in colors. Only a small part of the system is shown. We measured  $\bar{J}_V - \bar{J}_L = 0.04$ for $2\lambda=\zeta=0.5$, $\bar{J}_V - \bar{J}_L = 0.08$ for $2\lambda=\zeta=2.2$ and $\bar{J}_V - \bar{J}_L = -0.02$ for $2\lambda=\zeta=2$. This confirms that the instability arises only when the current on the liquid side overwhelms the one on the vapor side.}
    \label{fig:g}
\end{figure*}

\section{Profile $\varphi$ of the flat interface}	
As reported in the main text, the flat interfacial profile $\varphi(y)$ can be found analytically only for the case $2\lambda=\zeta$, where it equals the equilibrium case. For generic values of $\lambda,\zeta$, however, $w(\phi)=\varphi'^2$ can be obtained with a very simple numerical procedure using a technique introduced in~\cite{Wittkowski14} and then applied to AMB+ in~\cite{tjhung2018cluster}. We report it here for completeness. 

First, as shown in~\cite{tjhung2018cluster}, the binodals are easily obtained 
numerically solving $\mu = f'(\phi_1)=f'(\phi_2)$ and $\mu\psi_1-g(\phi_1) = \mu\psi_2-g(\phi_2)$. 
It is then easy to show that $w$ solves
\qq
K w' = (2\lambda-\zeta)w + 2(f'-\mu)\,
\qqq
which is solved by 
\qq\label{eq:app-interface-w-sol-gen}
w(x) = e^{-\frac{\zeta-2\lambda}{K} x} 
\left[ c + \frac{2}{K}\int_1^x  e^{\frac{\zeta-2\lambda}{K} y} (f'(x) -\mu) dy \right] .
\qqq
The knowledge of $\phi_{1,2}$ allows to fix the integration constant $c$ and the numerical evaluation of $w$ via (\ref{eq:app-interface-w-sol-gen}) is straightforward.



\section{Stability against normal perturbations}
We study here the linear stability to normal perturbations of the flat interface $\varphi$ at mean-field level ($D=0$). In this Appendix, for simplicity, we restrict to the case of one-dimensional interfaces. Due to mass conservation, the perturbed interface can be written as
\qq\label{app:eq:ansatz-1d-stability}
\phi(\bfx,y,t)=\varphi(y) + \p_y \epsilon(y,t)
\qqq
where $\varphi$ solves
\qq
\p_y^2 \left[
f''(\varphi)
-K \p_y^2\varphi 
+(\lambda-\zeta/2)\varphi'^2
\right]=0\,.
\qqq
Hence $\epsilon$ satisfies 
\qq
\p_t \epsilon = \mathcal{L}\epsilon+\mathcal{O}(\epsilon^2)
\qqq
where the linear operator  $\mathcal{L}$ is
\qq\label{app:eq-L}
\mathcal{L}
 = \p_y\left[ f''(\varphi)-K\p_y^2 +(2\lambda-\zeta)\varphi' \p_y\right]\p_y\,.
 \qqq
We are thus led to study the spectrum of $\mathcal{L}$. In the equilibrium case, and thus also when $\zeta=2\lambda$, it was shown in~\cite{shinozaki1993dispersion,bricmont1999stability} analytically that this is continuous for an infinite system and it touches $0$, resulting in algebraic decay of $\epsilon$ in time. This result relies on the fact that $\mathcal{L}$ is self-adjoint when $\zeta=2\lambda$. 
Extending this analysis to general $\zeta,\lambda$ lies beyond our scope. We compute numerically the spectrum of $\mathcal{L}$, concluding that the flat interface remains stable to normal perturbations irrespective of the sign of $\sigma$ and $\sigma_{\rm cw}$. Again the spectrum of $\mathcal{L}$ approaches $0$ in the large-system limit, indicating algebraic decay of $\epsilon$.

For generic $\lambda$ and $\zeta$, we studied numerically the spectrum of $\mathcal{L}$ for finite systems. By Fourier transforming along $y$, and for the choice of a double well local free energy $f$, 
 we consider the kernel $\mathcal{L}(q_1,q_2)$ of $\mathcal{L}$ defined from the relation $(\mathcal{L} \epsilon) (q_1)  = \int dq_2 \mathcal{L}(q_1,q_2) \epsilon(q_2)$ for any test function $\epsilon$. Explicitly:
\qq\label{eq:app:L-Fourier}
\mathcal{L}(q_1,q_2)
&=&
(-K q_1^4+A q^2)\delta(q_1-q_2)\\
&-&3A q_1 q_2 \mathcal{F}_y[\varphi^2](q_1-q_2)\nonumber\\
&+&(2\lambda-\zeta)q_1 q_2^2(q_1-q_2) \mathcal{F}_y[\varphi](q_1-q_2)\nonumber
\qqq
where $\mathcal{F}_y[\cdot]$ denotes the Fourier transform operator along $y$. We discretised $\mathcal{L}(q_1,q_2)$ on a grid with discretization step $\Delta x=1$ and total length $L_y$, so that $q_i=2\pi n_i/L_y$, $n_i=1,...,N$, $N\Delta x=L_y$. We then computed numerically the eigenvalues $\alpha_i$ of $\mathcal{L}$ for several values of $\lambda, \zeta$. Some of our results are reported in Fig. \ref{fig:spectrum}, showing that the qualitative picture is the same as at equilibrium: the spectrum of $\mathcal{L}$ is expected to be continuous and to touch $0$ for an infinite system. It should be observed that these conclusions apply irrespectively of the sign of both $\sigma$ and $\sigma_{cw}$: in both cases, $\varphi$ is stable against normal perturbations.

\begin{figure}[h!]\center
  \centering
    \includegraphics[width=1.\linewidth]{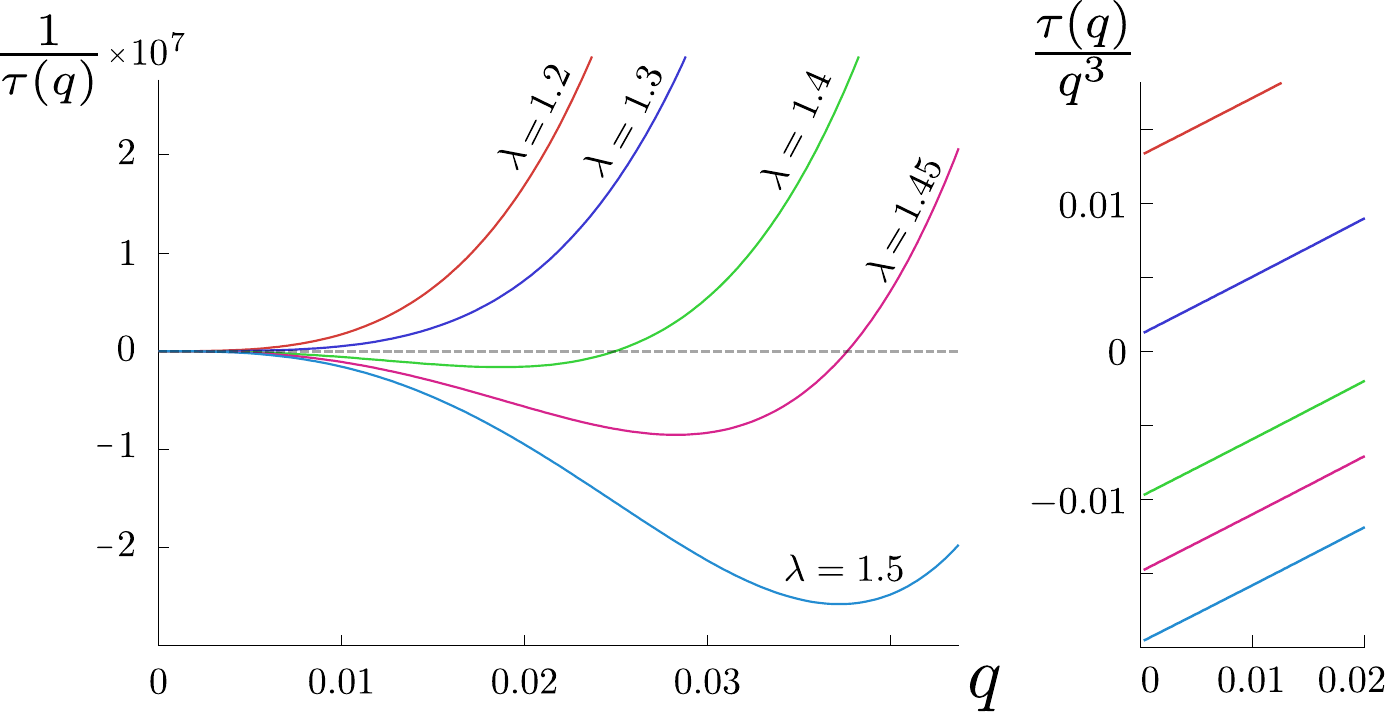}
    \caption{ (Left) Damping rate $1/\tau(q)$ vs $q$ at $\zeta=1$ crossing the stability line, located at $\lambda\simeq 1.29$. The most unstable mode (the minimum of $1/\tau(q)$) goes to $q=0$ as one approaches the critical $\lambda$. (Right) Plot of $\tau(q)/q^3$, showing that the change of sign of the damping rate happens at the estimated critical value of $\lambda$.}
    \label{fig:SM1}
\end{figure}


\section{Instability against height perturbations ($\sigma_{cw}<0$)}
In the main text, we have discussed the analogy between the instability of the flat interface taking place when $\sigma_{cw}<0$ and the Mullins-Sekerka instability~\cite{langer1980instabilities}. In Fig. \ref{fig:g} we further support this mechanistic picture, plotting the quasi-static current close to the perturbed interface. We consider three sets of parameter values corresponding to normal Ostwald ripening ($\sigma>0$), reversed Ostwald ripening but stable interface ($\sigma<0$, $\sigma_{cw}>0$), and unstable interface ($\sigma<0, \sigma_{cw}<0$). The current on the vapor side is always stabilizing while it is stabilizing in the liquid side only if $\sigma>0$. However, $\sigma<0$ is not sufficient to drive the instability. For this, the destabilizing current on the liquid side needs to be stronger than the one on the vapor side. This happens only in the rightmost case of Fig. \ref{fig:g}, which corresponds to $\sigma_{cw}<0$. To show this, we have measured the average current $\bar{J}_V$ in the liquid projected along $ e_\bfx=\bfx/|\bfx|$ defined as 
\qq
\bar J_L^2 = \int dx\int_{\{y|\phi(\bfx,y,t)>\phi_{th}\}} dy ({\bf J}\cdot e_\bfx)^2 
\qqq
and the analogous quantity in the vapor $\bar J_V$.

We further show that most unstable mode is the smallest one ($q=0$ in an infinite system) in the vicinity of the critical $\lambda$ value. To show this we leveraged on the fact that we have an analytic expression, eq. (6) of the main text, for the damping rate $1/\tau(q)$ in the effective interface equation. From simulations at $D=0$ and system-size $L_x=L_y=256$ we extracted the interfacial profile $\varphi(y)$ and then used it to evaluate $\tau(q)$ at arbitrarily low $q$. The results for $\zeta=1$ and varying $\lambda$ are reported in Fig. \ref{fig:SM1}.

\begin{figure}[h!]\center
  \centering
    \includegraphics[width=1.\linewidth]{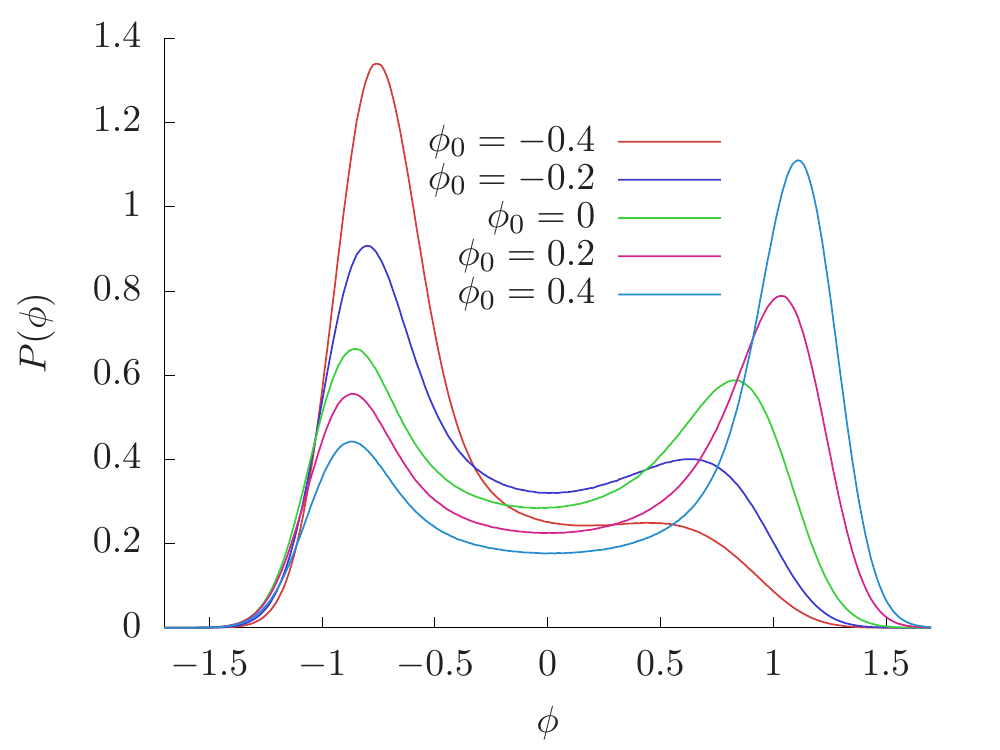}
    \caption{ PDF of the density for $\zeta=2$, $\lambda=1.75$, $D=0.05$ in a system $128\times 256$ and several global densities $\phi_0$ reported in the legend.}
    \label{fig:SM-PDF}
\end{figure}

\section{Liquid and vapor densities in the microphase separated and active foam states}
In Fig. 4, we report the PDF of the density 
as a function of the global density $\phi_0$. The vapor density is found, to a good accuracy, independent of $\phi_0$. Instead, the liquid density varies rather significantly with $\phi_0$. This is expected because of two reasons: the liquid density with which a finite size vapor bubble is in equilibrium differs from the binodal~\cite{tjhung2018cluster} and the presence of multiple droplets further change such value. Obtaining the dependence of the liquid density on $\phi_0$ is an open problem.

\begin{figure}[h!]\center
  \centering
    \includegraphics[width=0.8\linewidth]{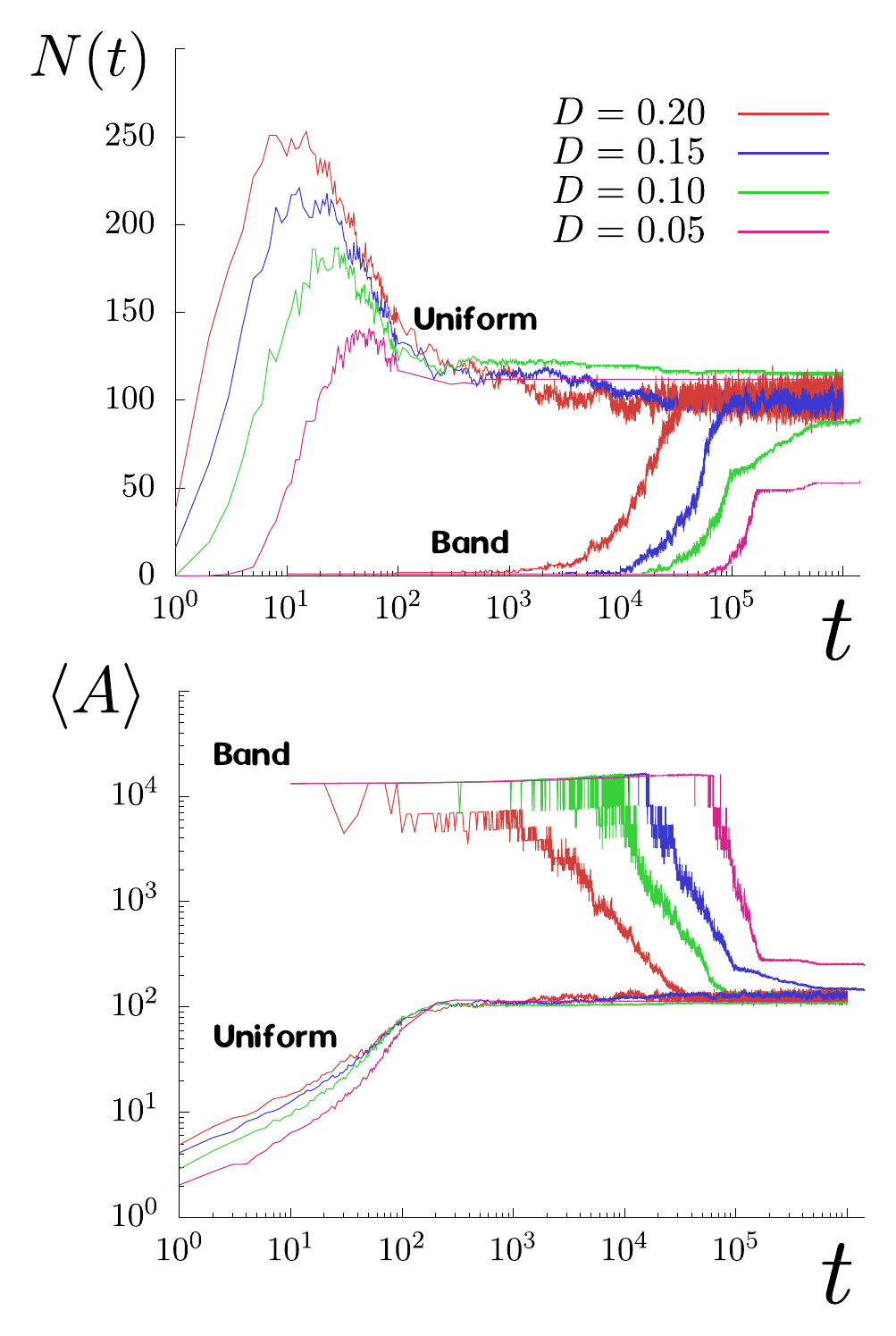}
    \caption{ Evolution in time of the average bubble size $\langle A\rangle$ and of their number $N(t)$ for $\zeta=2,\lambda=1.75$ and various noise values starting either from an uniform or from a band (a fully phase separated state with a flat interface).}
    \label{fig:evolution-MPS}
\end{figure}

\section{Evolution towards the microphase separated state}
In Fig. \ref{fig:evolution-MPS} we plot the evolution, starting from an homogeneous state or a fully phase separated state, 
 of the average size of bubbles and their number while converging to the microphase separated state. As shown, the convergence slows down when decreasing the noise value. This is because the initially formed bubbles are stable to small perturbations of their interface and evolution to the steady state is possible only by rare events at low noise. 

\section{Movies}
\begin{itemize}
\item Movie 1 : Interfacial instability starting from a fully phase separated initial condition with noise added in the bulk. Parameters: $D=0, \zeta=2.25, \lambda=1.8$. System size $L_x=L_y=128$. Total density $\phi_0=0.2$, leading to a microphase separation in the steady state. 

\item Movie 2 : Phase diagram for $D=0.05, \zeta=2,\lambda=1.75$ as a function of the density, showing the dynamics in the microphase separated and in the active foam state. System size $2L_x=L_y=256$.

\item Movie 3 : Evolution from a phase separated initial condition for $\zeta=1.5,\lambda=2.5, D=0.1$, which corresponds to $\sigma_{cw}<0$, and for $\zeta=2\lambda=2.2, D=0.25$, which corresponds to $\sigma_{cw}>0$. This show the markedly different evolution towards the microphase separated states found at $\sigma_{cw}<0$ and at $\sigma_{cw}>0$. Total simulation time is $10^6$ and system size $L_x=L_y=400$.
\end{itemize}

\end{document}